\documentclass[10pt, a4paper]{article}
\usepackage{lrec}
\usepackage{multibib}
\newcites{languageresource}{Language Resources}
\usepackage{graphicx}
\usepackage{tabularx}
\usepackage{soul}

\usepackage{epstopdf}
\usepackage[latin1]{inputenc}
\usepackage{balance}

\usepackage{soul}
\usepackage{color}

\definecolor{orangeX}{rgb}{1,0.5,0}
\definecolor{blueX}{rgb}{0.281, 0.359, 0.658}
\definecolor{purpleX}{rgb}{0.294118, 0, 0.509804}
\definecolor{greenX}{rgb}{0.721, 0.878, 0.341}
\graphicspath{{gfx/}} 
\newcommand{\eg}{e.\,g.\, }
\newcommand{\ie}{i.\,e.\, }

\usepackage[draft]{hyperref}
\usepackage{xstring}

\title{Responsible and Representative Multimodal Data Acquisition and Analysis: \\
On Auditability, Benchmarking, Confidence, Data-Reliance \& Explainability} 
\name{Alice Baird$^1$, Simone Hantke$^1$$^2$, Bj\"orn Schuller$^1$$^3$}

\address{$^1$ZD.B.Chair of Embedded Intelligence for Health Care and Wellbeing, University of Augsburg, Germany \\
$^2$Machine Intelligence and Signal Processing Group, Technische Universit\"at M\"unchen, Germany\\ 
$^3$GLAM -- Group on Language, Audio and Music, Imperial College London, UK \\
         \textit{alice.baird@informatik.uni-augsburg.de}\\}

\abstract{
The ethical decisions behind the acquisition and analysis of audio, video or physiological human data, harnessed for (deep) machine learning algorithms, is an increasing concern for the Artificial Intelligence (AI) community. 
In this regard, herein we highlight the growing need for responsible, and representative data collection and analysis, through a discussion of modality diversification. Factors such as Auditability, Benchmarking, Confidence, Data-reliance, and Explainability (ABCDE), have been touched upon within the machine learning community, and here we lay out these ABCDE sub-categories in relation to the acquisition and analysis of multimodal data, to weave through the high priority ethical concerns currently under discussion for AI. To this end, we propose how these five subcategories can be included in early planning of such acquisition paradigms. 
\\ 
\Keywords{Ethics, Multimodal, Representation, Data Acquisition,  Data Analysis, Machine Learning}
}

\begin{document}

\maketitleabstract

\begin{figure*}[t]
  \vspace{-0.1cm}
  \centering
  \includegraphics[trim={0.4cm 0.4cm 0.4cm 0.4cm},clip, width=0.6\linewidth]{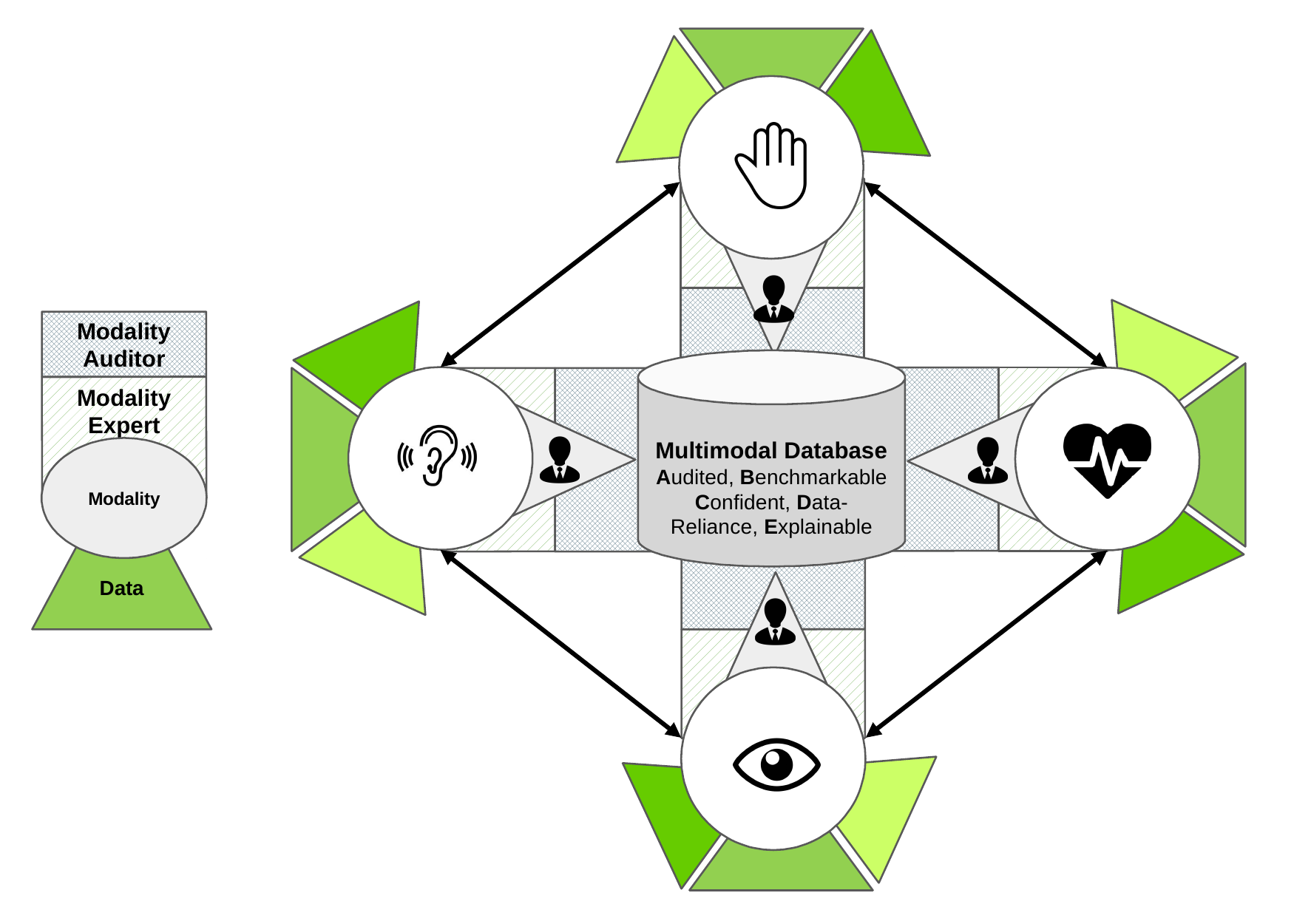}
  \vspace{-0.3cm}
  \caption{Proposed ABCDE approach for large-scale data collection across modalities. Considering multiple data sources, covering a spectrum of identities. Allowing for improved representation, as well as human safeguards within each modality.}
  \vspace{-0.4cm}
\end{figure*}

\section{Introduction}

Ethics itself is an encompassing term, 
and at its core it is the personal and societal decision making process between what is `right' and what is `wrong'. Although fundamental, 
within the research community the ethical (or moral) code of conduct can require quite some disentanglement, especially regarding Artificial Intelligence (AI)~\cite{AllenWS06}. Worldwide there are many non-technical groups of differing domains, which 
put ethics to the forefront of their manifesto, such as, the National Institutes of Health (NIH), or the National Science Foundation (NSF)~\cite{Resnik2007}. Additionally, some of the largest technology companies, such as Google's DeepMind are also bringing ethics to the forefront 
\footnote{https://deepmind.com/applied/deepmind-ethics-society/}, and multi-organisation ethics boards are being developed, such as the Partnership on AI \footnote{https://www.partnershiponai.org/board-of-directors/}.

Herein, when referring to ethics, we are namely focusing on the field of AI and in particular machine learning. The \textit{Ethics of AI}~\cite{Boddington17} has become a `hot topic' for researchers both internal and external to the AI community, with the peer-review process for publications itself being placed under scrutiny~\cite{PrecheltGF18}. 
The \textit{Ethics of AI} and \textit{Machine Ethics}~\cite{BaumHS18} are quite different terms for the field of AI. The latter refers to giving conscious ethical based decision making power to machines. The former although somewhat informing the latter, refers more broadly to decisions made by the researchers involved, covering diversity and representation, \eg to avoid discrimination~\cite{Zliobaite15a}, or inherent latent biases that may come from archival databases~\cite{abs-1801-01705}. As AI is largely data-driven, understanding its environment through context provided via annotations, 
\textit{Big Data} ethics for AI algorithms is an expanding discussion point~\cite{BerendtBR15,MittelstadtF16}. In this regard, crowdsourced data (\ie data gathered from large amounts of paid or unpaid individuals via the internet) -- made by humans with little time and minimal interest in the annotation task -- is one aspect which is raising concern, due to the resulting validity of such annotations~\cite{Hantke16-EFC}. 

Annotators are only limited by the classes provided to them by the researcher, and multi-modality during data labelling is only a technical challenge. A single domain model alone \eg a speech recognition system, can limit the potential `intelligence' of resulting AI algorithms, restricting the overall interaction. For truly `social' AI interactions, multimodal aware data should be of high priority, along with the pressing concerns of safety and privacy~\cite{Fox11}. Multimodal data, in turn brings more complex interpretation concerns \eg the term `grasp' is both a natural language processing, and gestural recognition problem~\cite{Mangin17}. Multiple modalities also multiplies the need for data auditability, and efficacy~\cite{Torresen18}. 

\section{Related Work and Motivations}
With AI efforts increasing, the ethical demands related to the needed Big Data, are expanding in parallel~\cite{herchel2017}. A general consideration which is not being completely overlooked, particularly in the field of Natural Language Processing~\cite{LeidnerP17}. With a vast amount of data sourced online, \eg through social media platforms, the legal aspects in terms of user privacy are at the forefront~\cite{wrro117179}. 

\textit{Machine Learning Fairness }(\ie bias), is currently a popular topic\footnote{https://developers.google.com/machine-learning/fairness-overview/}. With three core biases discussed \eg Interaction Bias, Latent Bias, and Selection Bias, in this paper we focus primarily on Selection Bias, as we propose that true multimodal and representative data could assist in avoiding this. Selection Bias \ie the by-products of decisions during collection and analysis -- including misrepresentation \eg through unbalanced gender classification~\cite{GaoA09a} -- is an important concern for technology companies. Such biasses can quite easily propagate into a resulting system, making it not only ethically problematic, but also fundamentally producing commercial limitations, \eg who does the model represent, and who will buy it?

Identity-representation itself is a prominent topic in AI today~\cite{Fussell2017}, and like many other human traits its manifestation within AI can be limited. The warnings are quite realistic, as bias created from the developers themselves~\cite{Zliobaite15a}, or through pre-existing archival (\ie data from historic sources) data~\cite{abs-1801-01705}, data-driven machine learning algorithms will simply replicate this. In this regard, the need for multimodal, representative data is more prevalent than ever. Not only, due to the aforementioned demographic biasses, including for gender-based variables~\cite{Larson17}, which can occur during collection, but also for representation and usability on a global scale, improving the overall impact of \textit{intelligent} HCIs. As well as this, for effective Human Computer Interactions (HCIs) social robotics require observation of a scene through multiple modalities by for example enhancing the ability for gesture recognition~\cite{PitsikalisKTM15}.  In the age of deep learning, multimodal interactions are the next stage for enhancing the usability of such algorithms~\cite{abs-1801-06048,BaltrusaitisAM17}. Offering benefits across domains, including conditions with a broad variety in population needs, such as Autism~\cite{LiuZZZL17}. 

\vspace{-0.15cm}
\section{Considerations}
With this motivation in mind, we now take a closer look at five core sub-categories which have been highlighted individually within the machine learning community, as factors which should be considered during data acquisitions and analysis, particularly as it pertains to multimodal data, as a means of informing a better identification representation within human data. Clearly, this represents only a very small sub-genre of the ethics within AI, yet of fundamental importance to the data-driven algorithm we quite commonly see, \eg embedded mobile assistants, or self-driving cars. These sub-categories include, Auditability, Benchmarking, Confidence, Data Reliance, and Explainability, and Figure 1 visualises a potential protocol, which can ensure that during collection and resulting analysis these factors can be considered in a more systematic way. 

\vspace{-0.15cm}
\subsection{Auditability} 
As the collection, annotation and analysis of truly Big Data can be extremely time consuming and costly for those involved, there are many methods now being developed to make collecting and annotating data in a semi-automatic fashion possible, including through social-media crawling~\cite{8273622}, and Active Learning annotation~\cite{Hantke17-TIC}. One consideration that engineers should keep in mind when designing such algorithms, is the balance between manual and automatic annotation. Having a human (manual) auditor present within such a database, is needed for many aspects from quality, to legality. Although the realisation may have technical challenges in the balance between public auditing and privacy~\cite{Diakopoulos2016}, the need for this has been highlighted in the literature in terms of security~\cite{WangWRLL11}, and privacy~\cite{TongSCL14}, with some showing concerns that autonomous agents responsible for collecting data, \eg directly from human manually labelled tags on YouTube, may breach privacy restriction or incorrectly store the resulting data due to an initial human error. In this regard, crowd-sourcing annotations should also be continually audited by human intervention with aspects such as fraud, being an ever present issue~\cite{RothwellECBRKK15}.

\vspace{-0.1cm}
\subsection{Benchmarking}
Benchmarking between system elements has shown to be difficult for multimodal, single domain databases~\cite{LiuXNSWK17}. A clear and explainable comparison between resulting systems has many advantages for development and improvement, yet aspects such as specific modalities are continually miss-matched across databases \ie in the image domain some may have RGB values, but miss the Skeleton extraction. In order to combine data, and fully utilise its potential, in the literature researchers have begun to closely look at benchmarking between multimodal databases~\cite{LiuFS11}, in this way improving HCI interactions. Through a more considered and ethical protocol during multimodal data collection, researchers could be guaranteed the possibility of benchmarking against the data of others.  

\vspace{-0.2cm}
\subsection{Confidence}  
Having `confidence' in one's database, comes across through these deeper considerations, and in turn offers in some way a more accurate long term system, with enhanced moral understanding~\cite{Blass18}. However, here we refer more to the use of confidence as a measure \ie how accurate is the current system prediction, as a means of understanding the current risk~\cite{Duncan2015}, additionally to aid future fault detection. This becomes specifically important when discussing AI systems which are designed for Human-Care, as not providing an overall confidence in data behind that, can result in a substantial risk to the human user~\cite{Ikuta2003}. Estimation of such attributes of data has been a focus for many researchers, with some also suggesting that the implementation of a `self-confidence' for social robotics could be a strong aid in avoiding accidents during interaction~\cite{Roehr2010}. Through multimodal near human-like perceptive data, confidence could be substantially improved across a system, as missing a single modality \eg audio domain, is an over all a weakness for the system, creating a blind-spot. 

\vspace{-0.1cm}
\subsection{Data-Reliance} 
Similar to confidence, data reliability is the process of completing acquisition of a database without error, within the context of the domain it is targeted towards~\cite{Morgan2004}. To show such reliability, there are standardised statistical tests such as p-values, which can be used as evidence of significance for particular aspects of the database. However the use of such tests, has begun to gain criticism in recent years, due to their extensive misuse by the machine learning community~\cite{VidgenY16}. Reliability of data is also discussed as a basis for the Internet of Things (IoT), as often (cf. Section 3.1) data is being sourced through social media (amongst other online sources) for the IoT, and researchers have suggested that such a single modal application may weaken the user experience. 

\vspace{-0.15cm}
\subsection{Explainabiltity} 
In recent years the machine learning community has been dominated by the need for accuracy, and a competitive nature has spread throughout the field, with internationally open challenges (asking participants to improve on a baseline system of a particular database), such as the Interspeech Computational Paralinguistics Challenges~\cite{SchullerSBVSRCWEMMSPVK13} and the ACM Audio / Visual Emotion Challenges~\cite{SchullerVECP12}, assisting in the rapid advancements of approaches across multiple domains. Although healthy competitions and high accuracy has its advantages, one limitation to this is the lack of explainabiltiy across the field~\cite{Huszar2015}. It is of up most importance that such machine learning models are interpretable, offering a clear use-case, as ultimately, without this, the results alone are potentially meaningless~\cite{VellidoML12}. As machine learning is predominately a pattern recognition task, visualisation of data has been a key enhancement for system explainabiltiy, particularly in deep learning, as this allows for trends within data to be more easily understood and interpreted~\cite{abs-1708-08296}. 

\vspace{-0.15cm}
\section{Concluding Remarks}
The current by-products of machine learning algorithms in relation to selection bias have been discussed throughout this paper, placing five key ABCDE sub-categories as they pertain to the enhancement of acquisition and analysis for multimodal data-driven models. We summarise these five key sub-categories which should be a considered focus to any researchers in the field of AI and machine learning. With ethics groups being founded by many of the largest AI based companies, there is clearly momentum towards ethical considerations within the community. One giant step forward in this regard is the multi-organisation discussion such as the Partnership on AI, but to the best of the authors' knowledge, there is not yet a large scale interdisciplinary discussion of researchers across domains. This would be a necessary step forward by the engineering community to include researchers from differing backgrounds, as AI is seemly going to be embedded in every aspect of our daily life, of which culture, and the arts, enrich substantially. In this regard, multi-modality should also not be restricted to the fundamental layers of the human senses, but expanded to more subtle differences, making way for a richer human imitation. This may be more appropriately achieved through interdisciplinary discussion, manifesting a more human-like tapestry of possibilities. Additionally in machine learning, the push toward \textit{0-shot}, or \textit{unsupervised learning} techniques, could be something to be wary of, and perhaps an audited, (at least semi-)supervised learning approach, is best for long-term implementations. With this in mind, considering a cross-modal data source, the implications of the five sub-categories discussed herein can be a step forward to a more representative AI model. 
\vspace{-0.15cm}
\section{Acknowledgements}
\noindent This work is funded by the Bavarian State Ministry of Education, Science and the Arts in the framework of the Centre Digitisation.Bavaria (ZD.B), and the European Union's Seventh Framework and Horizon 2020 Programmes under grant agreements No.\ 338164 (ERC StG iHEARu).







\footnotesize
\balance
\vspace{-0.15cm}
\section{Bibliographical References}
\bibliographystyle{output.bbl}
\bibliography{xample}

\end{document}